\documentclass[twocolumn,showpacs,amsmath,amssymb,superscriptaddress]{revtex4}
\input{psfig.sty}     
\begin{document}
\draft

\title{Magnetic-field dependence of electron spin relaxation in
n-type semiconductors}

\author{Franz X.~Bronold}
\affiliation{Institut f\"ur Physik,
Ernst-Moritz-Arndt Universit\"at Greifswald,
D-17487 Greifswald,  Germany}
\affiliation{Theoretical Division, Los Alamos National Laboratory,
Los Alamos, New Mexico 87545}
\author{Ivar Martin}
\affiliation{Theoretical Division, Los Alamos National Laboratory,
Los Alamos, New Mexico 87545}
\author{Avadh Saxena}
\affiliation{Theoretical Division, Los Alamos National Laboratory,
Los Alamos, New Mexico 87545}
\author{Darryl L. Smith}
\affiliation{Theoretical Division, Los Alamos National Laboratory,
Los Alamos, New Mexico 87545}
\date{\today}  

\begin{abstract}
{We present a theoretical investigation of the magnetic field
dependence of the longitudinal ($T_1$) and transverse ($T_2$) spin
relaxation times of conduction band electrons in n-type III-V
semiconductors. 
In particular, we find that the
interplay between the Dyakonov-Perel process and an additional
spin relaxation channel, which originates from the electron 
wave vector dependence of the electron $g$-factor,
yields a maximal $T_2$ at a finite
magnetic field. We compare our results with existing experimental
data on n-type GaAs and make specific additional predictions for
the magnetic field dependence of electron spin lifetimes.}
\end{abstract}


\maketitle

The electron spin in a semiconductor is a robust object which can
be utilized to add new functionality to existing electronic
devices or to even build completely new devices based on this spin
degree of freedom.~\cite{Wolf01}  Establishment of successful
spintronics devices requires a thorough understanding of the
electron spin dynamics in a semiconducting environment. In
particular, spin relaxation processes need to be identified and
controlled.

Important electron spin relaxation processes in n-type
semiconductors include the Elliott-Yafet
(EY) process~\cite{Elliott54,Yafet63}, that leads to spin flip
scattering and, in semiconductors without inversion symmetry, the
Dyakonov-Perel (DP) process~\cite{Dyakonov72}, in which spin
states precess because of spin off-diagonal Hamiltonian matrix
elements resulting from a combination of the spin-orbit
interaction and inversion asymmetry. Typically, the DP mechanism
dominates the spin dynamics in n-type III-V semiconductors. An
external magnetic field, in many cases required for control and
manipulation in spintronics devices, can significantly influence
electron spin dynamics. A magnetic field has two main effects on
electron spin relaxation: (i) it quenches the DP process thereby
tending to extend the spin lifetimes as a function of the magnetic
field~\cite{Ivchenko73}, and (ii) it opens an additional spin
relaxation process which tends to reduce the spin lifetimes in
applied magnetic fields.~\cite{Margulis83} The latter process is
due to the wave vector dependence of the conduction band (CB)
electron $g$-factor. As a result of the variations in the
$g$-factor, electrons in different quantum states precess about a
transverse magnetic field at different rates and thus lose spin
coherence. For brevity we will refer to this process as a variable
$g$-factor (VG) mechanism.

In contrast to previous 
studies~\cite{Elliott54,Yafet63,Dyakonov72,Ivchenko73,Margulis83,Chazalviel75,Fishman77,Wu00,
Lau01,Song02}
of spin relaxation in (bulk)
n-type III-V semiconductors, we simultaneously treat
the EY, DP,
and VG processes on an equal footing and focus on the interplay
between the various spin relaxation processes as a function of the
magnetic field. Thereby, we are able to study in detail the competition 
between the quenching of the DP process and the appearance of the 
VG process. 

Specifically, we calculate 
the longitudinal ($T_1$)
and transverse ($T_2$) spin relaxation times as a function of 
temperature, electron density and magnetic field.  
We find that the VG process dominantly
influences the transverse ($T_2$) spin relaxation time. In
particular, as a result of the competition between the quenching
of the DP process and the introduction of the VG process, there is
a magnetic field for which the transverse ($T_2$) spin lifetime is
maximal. From the slope of $T_2$ at small magnetic fields it is
moreover possible to determine whether the DP or the EY process
dominates spin relaxation at zero magnetic field. In contrast, the
magnetic field dependence of the longitudinal ($T_1$) lifetime is
essentially unaffected by the VG process and dominated by the
quenching of the DP process. Thus, they generally increase with  
field and saturate at a value given by the EY process.

In an applied magnetic field, the CB electrons in a III-V
semiconductor, e.g. GaAs, are described by the
Hamiltonian~\cite{Ogg66,Golubev85}
\begin{eqnarray}
  {\cal H}_{\alpha\beta}(\vec{K})=
  \epsilon(\vec{K})\delta_{\alpha\beta}
  +{\hbar\over 2}
  \large[\vec{\Omega}_L+\vec{\Omega}_{IA}(\vec{K})+
  \vec{\Omega}_g(\vec{K})\large]
  \!\cdot\!\vec{\sigma}_{\alpha\beta} ,
  \label{model}
\end{eqnarray}
where $\vec{K}=\vec{k}-(e/\hbar c)\vec{A}(\vec{r})$ [$\vec{A}(\vec{r})$
is the vector potential], $\epsilon(\vec{k})$ is the Kramers
degenerate dispersion of CB electrons,
$\hbar\vec{\Omega}_L=\mu_B g^*\vec{B}$ is the Larmor frequency,
$\hbar\vec{\Omega}_{IA}(\vec{K})=2\delta_0\vec{\kappa}(\vec{K})$
is the splitting of the CB dispersion due to the combination of
spin-orbit interaction and inversion asymmetry, and
\begin{eqnarray}
  \hbar\vec{\Omega}_g(\vec{K})=2 a_4 K^2 \vec{B}
  + 2 a_5\{\vec{K},\vec{B}\!\cdot\!\vec{K}\}
  + 2 a_6\vec{\tau}(\vec{K},\vec{B})
  \label{Omegag}
\end{eqnarray}
is a term which gives rise to a wave vector dependence of the CB
electron $g$-factor. The definitions of the vectors
$\vec{\kappa}(\vec{K})$ and $\vec{\tau}(\vec{K},\vec{B})$ and of
the parameters $\delta_0$, $a_i, i=4,5,6$ are given in
Refs.~\cite{Ogg66,Golubev85} and $\{,\}$ indicates an
anticommutator.

Our calculation starts from the full quantum kinetic equations for the
contour-ordered Green functions~\cite{Kinetic}, from which we derive, 
considering a classical homogeneous magnetic field and using
the fact that wave vector scattering is essentially instantaneous
on the time scale of spin relaxation, a semiclassical kinetic
equation for the CB electron density matrix. We then linearize this
kinetic equation with respect to the CB electron spin density, 
assuming, as an initial condition, small spin polarization.
Treating scattering processes in the Born approximation and
expanding the collision integrals up to second order in the wave
vector transfer (diffusion approximation~\cite{Kinetic}), we
finally obtain a generalized Fokker-Planck-Landau equation for the spin
density $\vec{S}(\vec{k}t)$ which, in atomic units (with magnetic field
along the z-axis) reads
\begin{widetext}
  \begin{eqnarray}
    \large[\partial_t-i\Omega_C\hat{\cal L}_z~\large]
    \vec{S}(\vec{k}t)=
    \large[\vec{\Omega}_L+\vec{\Omega}_{IA}(\vec{k})+
    \vec{\Omega}_g(\vec{k})\large]\times\vec{S}(\vec{k}t)
    +\large[\hat{\cal D}-{1\over{2\tau_1(k)}}{\hat{\cal L}^2}\large]
     \vec{S}(\vec{k}t)
    -\Gamma(\vec{k})\vec{S}(\vec{k}t),
    \label{FPL}
  \end{eqnarray}
\end{widetext}
with $\Omega_C$ the cyclotron frequency, $\hat{\cal L}_z$ and
$\hat{\cal L}^2$ the z-component and the squared total angular
momentum operator in wave vector space, respectively, $1/\tau_1(k)$
the sum of the (on-shell) wave vector relaxation rates for the various
scattering processes, and $\hat{\cal D}$ a differential operator in
$k=|\vec{k}|$ relevant to inelastic scattering processes.

Equation (\ref{FPL}) contains EY, DP, and VG processes and
accounts for Larmor precession and orbital motion of the 
CB electrons in the magnetic field. More specifically, 
the EY process, due to genuine spin flip scattering events, is 
given by the tensor $\Gamma$, whereas the DP and VG processes 
originate from the interplay of spin conserving 
wave vector scattering events described by the differential operator 
$\hat{\cal D}-(1/2\tau_1){\hat{\cal L}^2}$ and the torque forces 
due to $\vec{\Omega}_{IA}$ and $\vec{\Omega}_g$, respectively. 
The orbital motion encoded in $-i\Omega_C\hat{\cal L}_z$ and, to a 
lesser extend, the torque force due to $\vec{\Omega}_L$ lead to a 
quenching of the DP process.

It is possible to derive from Eq. (\ref{FPL}) general expressions 
for the spin relaxation rates without specifying whether the scattering 
processes are elastic ($\hat{\cal D}=0$) or inelastic ($\hat{\cal D}\neq 0$).
To that end, we follow 
Ref.~\cite{Dyakonov72} and employ a perturbative approach with
respect to the torque forces. Our results are therefore valid for
$|\vec{\Omega}_{IA}+\vec{\Omega}_g|\tau_1 < 1$. Expanding
$\vec{\Omega}_{IA}(\vec{k})$ and $\vec{\Omega}_{g}(\vec{k})$ in
terms of spherical harmonics $Y_{lm}(\Theta,\Phi)$, we 
find ($i=1,2$)~\cite{Bronold02}
\begin{eqnarray}
  \left[T_i\right]^{-1}=\left[T_i^{EY}\right]^{-1}+
  \left[T_i^{DP}\right]^{-1}+\left[T_i^{VG}\right]^{-1},
  \label{Ti}
\end{eqnarray}
with the EY contributions (due to spin-flip scattering)
\begin{eqnarray}
  \left[T_1^{EY}\right]^{-1}=2\left[T_2^{EY}\right]^{-1}
                  ={{32\pi}\over3} C_{sf}^2
                     \left\langle{{k^4}\over{\tau_1(k)}}\right\rangle,
\label{EY}
\end{eqnarray}
the DP contributions (due to inversion asymmetry)
\begin{eqnarray}
  \left[T_1^{DP}\right]^{-1}\!\!&=&\!4 |C_{31}^Y|^2
        \tilde{\tau}_{31}^3
        +4 |C_{33}^Y|^2
        \tilde{\tau}_{33}^3,
  \\
  \left[T_2^{DP}\right]^{-1}\!\!&=&\!2 |C_{32}^Z|^2 \tilde{\tau}_{32}^3
  +2 |C_{31}^Y|^2\tilde{\tau}_{31}^3
  +2 |C_{33}^Y|^2\tilde{\tau}_{33}^3,
  \label{DP}
\end{eqnarray}
and the VG contributions (due to the wave vector dependence of the CB
electron $g$-factor)
\begin{eqnarray}
  \left[T_1^{VG}\right]^{-1}\!\!&=&\!4 |D_{21}^Y|^2\tilde{\tau}_{21}^1,
  \label{T1VG}\\
  \left[T_2^{VG}\right]^{-1}\!&=&\!\!|D_{00}^Z|^2\tilde{\tau}_{00}^1
  +|D_{20}^Z|^2\tilde{\tau}_{20}^1
  +2 |D_{21}^Y|^2\tilde{\tau}_{21}^3.
  \label{T2VG}
\end{eqnarray}
Here, $C_{lm}^i$ and $D_{lm}^i$ ($i=X, Y, Z$) are the expansion coefficients
of ${\Omega}^i_{IA}(\vec{k})$ and ${\Omega}^i_{g}(\vec{k})$,
respectively, and ($\nu=1, 2, 3$)
\begin{eqnarray}
\tilde{\tau}_{lm}^\nu=-\mbox{\rm Re}\langle C_l(k)\tau_{lm}^\nu(k)\rangle,
\label{taulm}
\end{eqnarray}
where the brackets denote an average over $k$ defined as 
$\langle~(~...~)~\rangle=
\int_0^\infty dk k^2 f(k)\bar{f}(k)(...)/4\pi\int_0^\infty dk k^2
f(k)\bar{f}(k)$, with $\bar{f}(k)=1-f(k)$ and $f(k)$ the equilibrium
Fermi distribution function. The 
generalized wave vector relaxation time
$\tau_{lm}^\nu(k)$ satisfies a differential equation
\begin{eqnarray}
\large[\hat{\cal D}+i(m\Omega_C+\Omega_L^\nu)-{1\over{\tau_l(k)}}
\large]\tau_{lm}^\nu(k)=C_l(k),
\label{diffeqn}
\end{eqnarray}
with $C_l(k)=C_{IA}k^3\delta_{l3}+C_g(B)k^2[\delta_{l2}+\delta_{l0}]$,
$\Omega_L^\nu=\Omega_L\large[\delta_{\nu1}-\delta_{\nu2}\large]$, and
$1/\tau_l(k)=l(l+1)/2\tau_1(k)$.
The constants characterizing the three spin relaxation processes
are, respectively,
$C_{sf}=\delta^2(\Delta+2\epsilon_g)R_0m_0/2\Delta\epsilon_gm^*$,
$C_{IA}=2\delta_0/R_0a_0^3$, and $C_g(B)=2\mu_B B/R_0 a_0^2$, where
$\delta^2=2\Delta^2/(\Delta+\epsilon_g)(2\Delta+3\epsilon_g)$, 
$\Delta$ is the spin-orbit splitting, $\epsilon_g$ is the band
gap, $R_0$ and $a_0$ are the Rydberg energy and the Bohr
radius, respectively, $m^*$ and $m_0$ are the CB electron mass
and the mass of a bare electron, respectively, and $\mu_B$ is the Bohr
magenton. The detailed form of $\hat{\cal D}$ depends on the scattering
processes and does not concern us here. \cite{Bronold02} 

Note, as a consequence of the orthogonality of the angle dependences, 
the EY, DP, and VG spin relaxation rates are additive. The 
generalized relaxation rate $1/\tau_{lm}^\nu(k)$, on the other hand,
is in general not proportional to the sum of the (on-shell) 
relaxation rates  $1/\tau_l(k)$ because of 
inelasticity. A Matthiessen-type rule for $1/\tau_{lm}^\nu(k)$ only holds 
for elastic scattering (see below).

We are interested in the magnetic field dependence of the
spin relaxation processes which, at least qualitatively, 
should not depend on 
the approximation adopted to decribe the scattering events. 
In the following, we treat therefore all scattering processes in
the elastic approximation and neglect $\hat{\cal D}$ in
Eq. (\ref{diffeqn}). Specifically, we take scattering on 
ionized impurities, acoustic phonons, and longitudinal optical (LO)
phonons into account. The elastic approximation restricts our results to
low enough temperatures, where electron-impurity scattering
dominates, and to high enough temperatures, where electron-phonon
scattering becomes essentially elastic.   

Within the elastic approximation Eq. (\ref{diffeqn}) reduces to 
an algebraic equation which is readily solved to yield
\begin{eqnarray}
  \tilde{\tau}_{lm}^\nu=
  \left\langle
  {{\tau_l(k)\left[C_{IA}^2 k^6\delta_{l3}+C_g^2(B)k^4 (\delta_{l2}+\delta_{l0})\right]}
  \over{1+[(m\Omega_C+\Omega_L^\nu)\tau_l(k)]^2}}
  \right\rangle.
\end{eqnarray}
The k-average can be obtained either numerically or, at low and high temperatures,
with saddle point techniques exploiting the
peaked structure of the integrands. Within the elastic approximation 
it is sufficient to adopt the latter. Details of the calculation 
will be given elsewhere.~\cite{Bronold02} 
\begin{figure}[t]
\psfig{figure=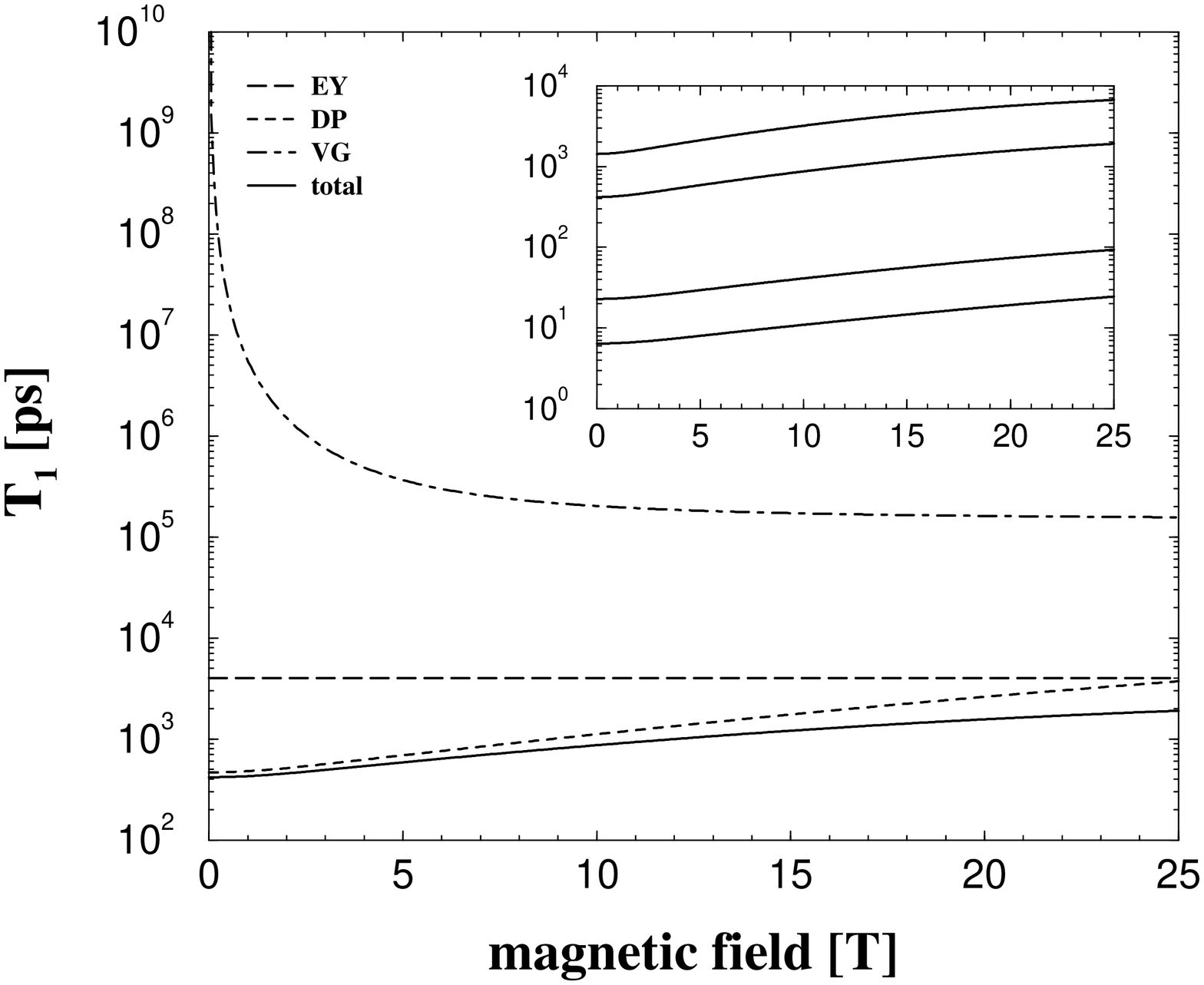,height=6.0cm,width=0.95\linewidth,angle=-90}
\psfig{figure=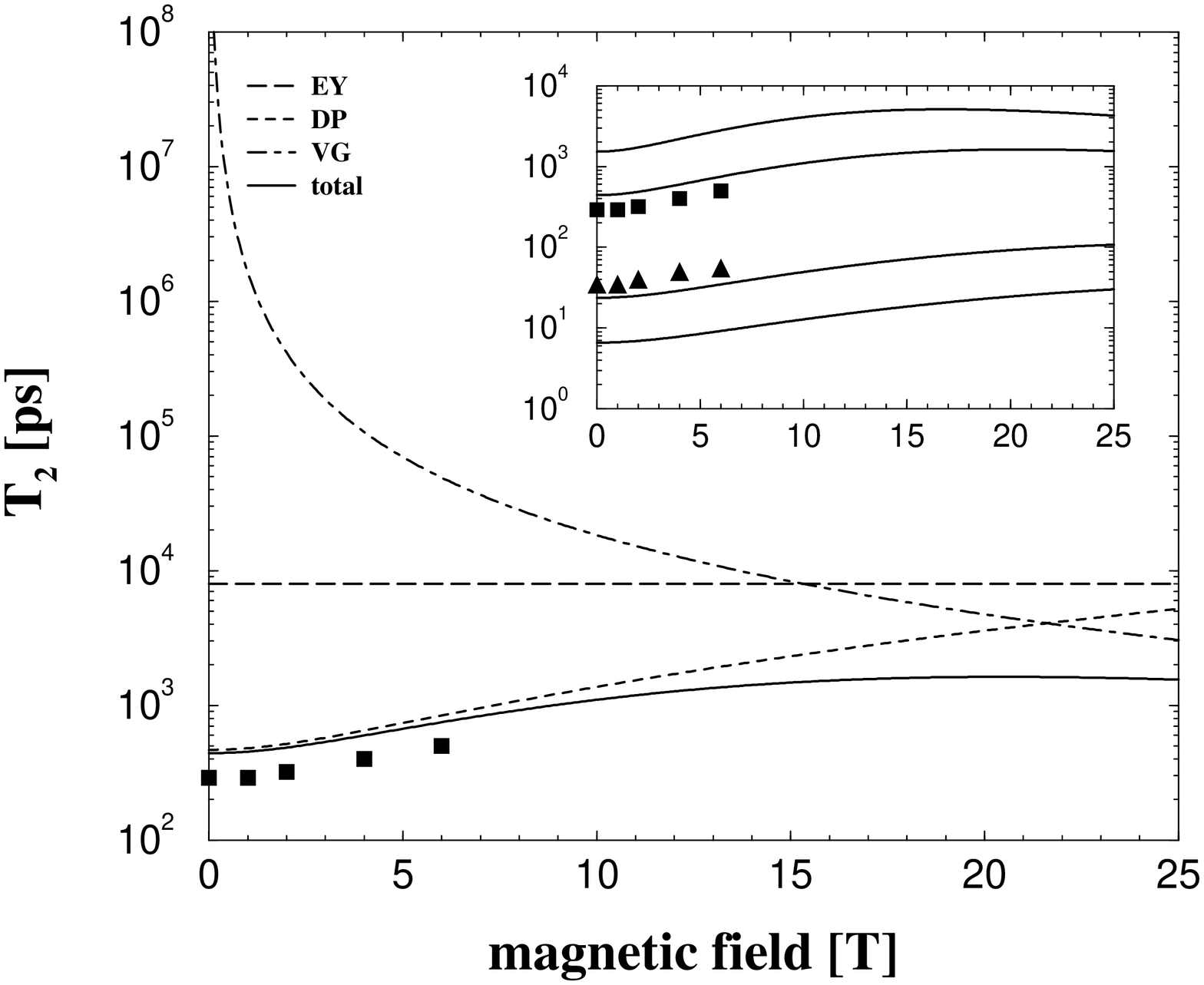,height=6.0cm,width=0.95\linewidth,angle=-90}
\caption[fig1] {The top and bottom panels show, respectively,
$T_1$ and $T_2$ in GaAs as a function of magnetic field for $T=0$
and $n=10^{18}\mbox{cm}^{-3}$.  The contributions from the EY (long
dash), DP (short dash) and VG (dot-dash) processes and the total
relaxation time (solid) are shown in the main panel. The insets
(same axis as the main panel)
show the total relaxation times for 
$n=5\times 10^{17}\mbox{cm}^{-3}, 1\times
10^{18}\mbox{cm}^{-3}, 5\times 10^{18}\mbox{cm}^{-3},$ and $1\times
10^{19}\mbox{cm}^{-3}$ (top to bottom). The squares and triangles are
experimental data from Ref.\cite{Kikkawa98} at the respective densities.}
\label{fig1}
\end{figure}   

\begin{figure}[t]
\psfig{figure=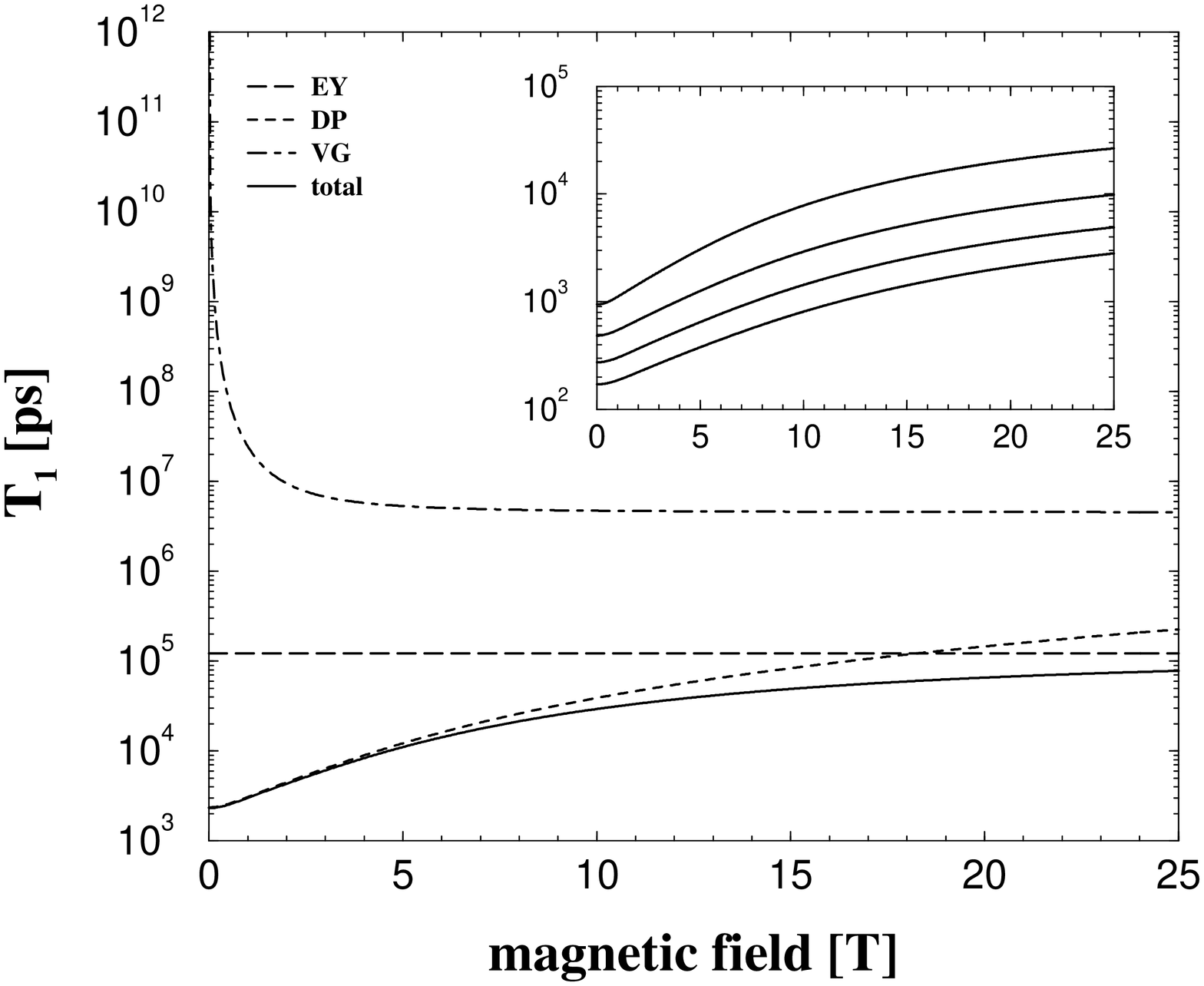,height=6.0cm,width=0.95\linewidth,angle=-90}
\psfig{figure=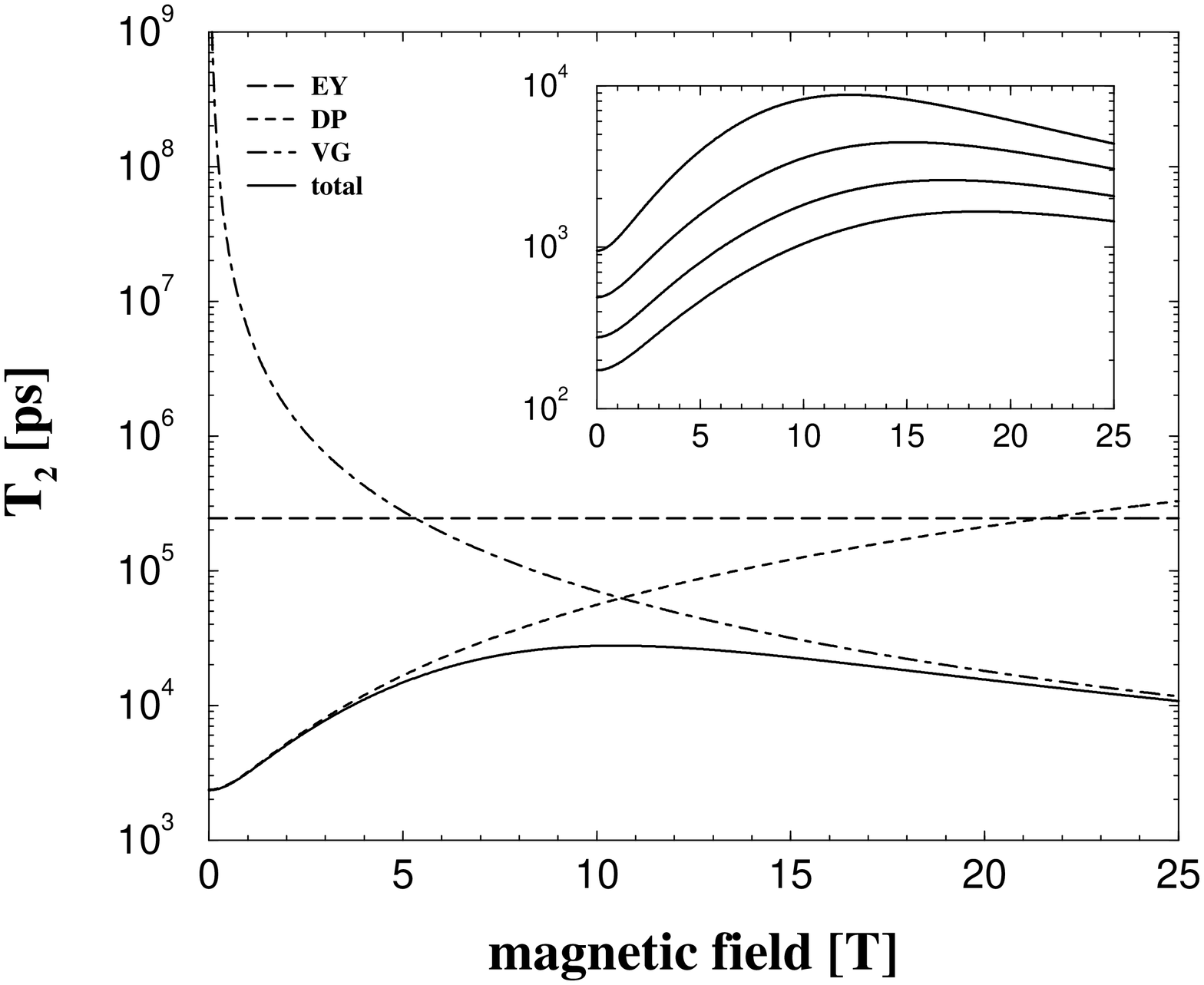,height=6.0cm,width=0.95\linewidth,angle=-90}
\caption[fig2] {The top and bottom panels show, respectively, $T_1$
and $T_2$ in GaAs as a function of magnetic field for $T=100\mbox{K}$ and
$n=10^{17}\mbox{cm}^{-3}$. The contributions from the EY (long dash), DP
(short dash) and VG (dot-dash) processes and the total relaxation
time (solid) are shown in the main panel.  The insets 
(same axis as the main panel) show the
total relaxation times for $T=150\mbox{K}, 200\mbox{K}, 
250\mbox{K},$ and $300\mbox{K}$ (top
to bottom). } \label{fig2}
\end{figure}        

In Fig. \ref{fig1}, we show calculated longitudinal ($T_1$) and
transverse ($T_2$) spin relaxation times for GaAs as a function of
magnetic field at $T=0$ and an electron density of
$n=10^{18}\mbox{cm}^{-3}$. We show separately the contributions to the
spin relaxation times from the EY, DP and VG processes and the
total spin relaxation time including all three spin relaxation
processes. In the insets of Fig. \ref{fig1}, we give the total
spin relaxation time for various electron densities at $T=0$.  
The parameters needed to specify $\vec{\Omega}_g(\vec{k})$
have been previously obtained partly experimentally by measuring
combined cyclotron resonances ($a_4, a_6$)
and partly theoretically within a
five-level Kane model ($a_5$):
$(a_4, a_5, a_6)=(97,-8,49)\times 10^{-24}
\mbox{eVcm}^2\mbox{Oe}^{-1}$.~\cite{Golubev85}
The parameter defining $\vec{\Omega}_{IA}(\vec{k})$ is given by
$\delta_0=0.06\hbar^3/\sqrt{(2m^*)^3\epsilon_g}$.~\cite{Aronov83}
The remaining parameters,
such as the effective CB electron mass or the deformation
potential are available from standard data bases.~\cite{parameter}

For the temperature and density conditions in Fig. \ref{fig1},
the electrons are degenerate and electron-ionized impurity
scattering dominates. The VG process makes a small contribution
to $T_1$ which is dominated by the DP process at zero magnetic
field. As the magnetic field is increased, the DP process is
quenched.
Thus, $T_1$ increases monotonically
with increasing magnetic field saturating at high field at a value
determined by the EY process which is not affected by the
magnetic field.~\cite{comment} If the material parameters had been
such that the EY process dominated the DP process for $T_1$
relaxation at zero magnetic field, $T_1$ relaxation would not be
significantly affected by the applied field.
By contrast the VG
process makes a significant contribution to $T_2$ relaxation.  At
small applied magnetic fields the $T_2$ lifetime increases with
increasing magnetic field, but as the field continues to increase
the VG process begins to dominate the relaxation so that $T_2$ has
a maximum and begins to decrease for larger magnetic fields.  If
the material parameters had been such that the EY process
dominated the DP process for $T_2$ relaxation at zero magnetic
field, the $T_2$ relaxation would monotonically decrease with
increasing magnetic field.

The solid squares and triangles in the lower panel of Fig. \ref{fig1}
are measured $T_2$ spin lifetimes
in GaAs at 5K from
Ref. \cite{Kikkawa98} at electron densities of $1\times
10^{18}\mbox{cm}^{-3}$ and $5\times 10^{18}\mbox{cm}^{-3}$.
(Data for an electron concentration of $1\times 10^{16}\mbox{cm}^{-3}$ at
5K was also presented in Ref. \cite{Kikkawa98}, but at this low density the
electrons are bound to isolated donors and our theory does not apply.)
There is good (order of magnitude) agreement between our calculation
and these measured results, although there were no adjustable
parameters. Unfortunately, the magnetic fields in Ref. \cite{Kikkawa98}
are not high enough to capture any effects due to the VG process.
In particular, our prediction of the maximum of $T_2$ remains to be experimentally
verified.

In Fig. \ref{fig2}, we show the various contributions to the 
$T_1$ and $T_2$ spin relaxation  
for GaAs as a function of magnetic field at
$T=100\mbox{K}$ and an electron density of $n=10^{17}\mbox{cm}^{-3}$. In the
insets of Fig. \ref{fig2}, we show the total spin relaxation time as a
function of magnetic field for various temperatures at
$n=10^{17}\mbox{cm}^{-3}$.
For the temperature and density conditions in Fig. \ref{fig2}, the
electrons are non-degenerate and electron-LO-phonon scattering is
the dominant scattering process. As for the degenerate electron
case, the VG process makes a small contribution to $T_1$ which is
again dominated by the DP process at zero magnetic field.
The DP process is quenched by the field so that $T_1$ increases with
field at small fields and saturates at a value determined by the
EY process at large fields.~\cite{comment}
Similar to the degenerate electron 
case, the VG process makes a
substantial contribution to $T_2$ relaxation. At small fields
the $T_2$ lifetime increases with increasing field and at large
fields the VG process begins to dominate the relaxation so that
$T_2$ has again a maximum at some finite magnetic field. 
The sign of the slope in $T_2$ at small magnetic
fields is again a clear signature of whether the EY process ($T_2$
decreases with increasing field) or DP process ($T_2$ increases
with increasing field) dominates $T_2$ relaxation at zero magnetic
field.           
Note, the qualitative behavior of the longitudinal
and transverse spin relaxation times with increasing magnetic
field is similar for degenerate and non-degenerate electrons, but
the magnitude of the change is larger for non-degenerate
electrons. 

In summary, based on a systematic kinetic approach, which treats
the EY, DP, and VG processes on an equal footing, we calculated
the longitudinal ($T_1$) and transverse ($T_2$) spin relaxation
times of CB electrons in n-type III-V semiconductors as a function
of temperature, electron density, and magnetic field. At finite
magnetic field, the VG process competes with the DP and EY
processes. We find that, as a consequence of the interplay of the
DP and the VG processes, $T_2$ can have a maximum as a function of
magnetic field. In contrast, $T_1$ is not affected by the VG
process and increases with magnetic field until it saturates at a
value determined by the EY process.
The sign of the change in
$T_2$ with increasing magnetic field at small fields indicates,
moreover, whether the EY process or
the DP process dominates
$T_2$ relaxation at zero magnetic field. Our calculated results
are in good agreement with existing experimental data in n-type
GaAs and we make additional specific predictions for the magnetic
field dependence of electron spin lifetimes that are subject to
experimental check.

This work was supported by the SPINs program of
the U.S. Defense Advanced Research Projects Agency.



\begin{references}
\bibitem{comment}
This contradicts the results given in Ref.~\cite{Margulis83}. 
The discrepancy has to be due to the too  
large $a_5$ value used in Ref.~\cite{Margulis83}.
\end{references}
\end{document}